\documentclass[twocolumn,twocolappendix,times,tighten]{aastex631}

\usepackage{multirow}

\newcommand{\kms}{\mbox{km s$^{-1}$}}                                    

\published{in ApJL}

\shorttitle{ALMA Detection of a PMC Disk}
\shortauthors{Wu et al.}

\begin{document}
\title{{\large ALMA Discovery of a Disk around the Planetary-mass Companion SR 12 c}}

\author{Ya-Lin Wu}
\affiliation{Department of Physics, National Taiwan Normal University, Taipei 116, Taiwan; {\rm yalinwu@ntnu.edu.tw}}
\affiliation{Center of Astronomy and Gravitation, National Taiwan Normal University, Taipei 116, Taiwan}

\author{Brendan P. Bowler}
\affiliation{Department of Astronomy, The University of Texas at Austin, Austin, TX 78712, USA}

\author{Patrick D. Sheehan}
\altaffiliation{NSF Astronomy \& Astrophysics Fellow}
\affiliation{Department of Physics and Astronomy, Northwestern University, 2145 Sheridan Road, Evanston, IL 60208, USA}

\author{Laird M. Close}
\affiliation{Steward Observatory, University of Arizona, 933 North Cherry Avenue, Tucson, AZ 85721, USA}

\author{Joshua A. Eisner}
\affiliation{Steward Observatory, University of Arizona, 933 North Cherry Avenue, Tucson, AZ 85721, USA}

\author{William M. J. Best}
\affiliation{Department of Astronomy, The University of Texas at Austin, Austin, TX 78712, USA}

\author{Kimberly Ward-Duong}
\affiliation{Smith College, 10 Elm Street, Northampton, MA 01063, USA}

\author{Zhaohuan Zhu}
\affiliation{Department of Physics and Astronomy, University of Nevada, Las Vegas, 4505 S. Maryland Pkwy, Las Vegas, NV 89154, USA}

\author{Adam L. Kraus}
\affiliation{Department of Astronomy, The University of Texas at Austin, Austin, TX 78712, USA}

\begin{abstract} 
\noindent We report an Atacama Large Millimeter/submillimeter Array 0.88 mm (Band 7) continuum detection of the accretion disk around SR 12 c, an $\sim$11 $M_{\rm Jup}$ planetary-mass companion (PMC) orbiting its host binary at 980 au. This is the first submillimeter detection of a circumplanetary disk around a wide PMC. The disk has a flux density of $127 \pm14~\mu$Jy and is not resolved by the $\sim$0\farcs1 beam, so the dust disk radius is likely less than 5 au and can be much smaller if the dust continuum is optically thick. If, however, the dust emission is optically thin, then the SR 12 c disk has a comparable dust mass to the circumplanetary disk around PDS 70 c but is about five times lower than that of the $\sim$12 $M_{\rm Jup}$ free-floating OTS 44. This suggests that disks around bound and unbound planetary-mass objects can span a wide range of masses. The gas mass estimated with an accretion rate of $10^{-11}~M_\sun$ yr$^{-1}$ implies a gas-to-dust ratio higher than 100. If cloud absorption is not significant, a nondetection of ${}^{12}$CO(3--2) implies a compact gas disk around SR 12 c. Future sensitive observations may detect more PMC disks at 0.88 mm flux densities of $\lesssim$100 $\mu$Jy.\\ \\
{\it Unified Astronomy Thesaurus concepts:} Radio continuum emission (1340); Radio interferometry (1346); Accretion (14); Extrasolar gaseous giant planets (509)\\
\end{abstract}

\section{Introduction}

\begin{figure*}[t]
\centering
\figurenum{1}
\includegraphics[width=\linewidth]{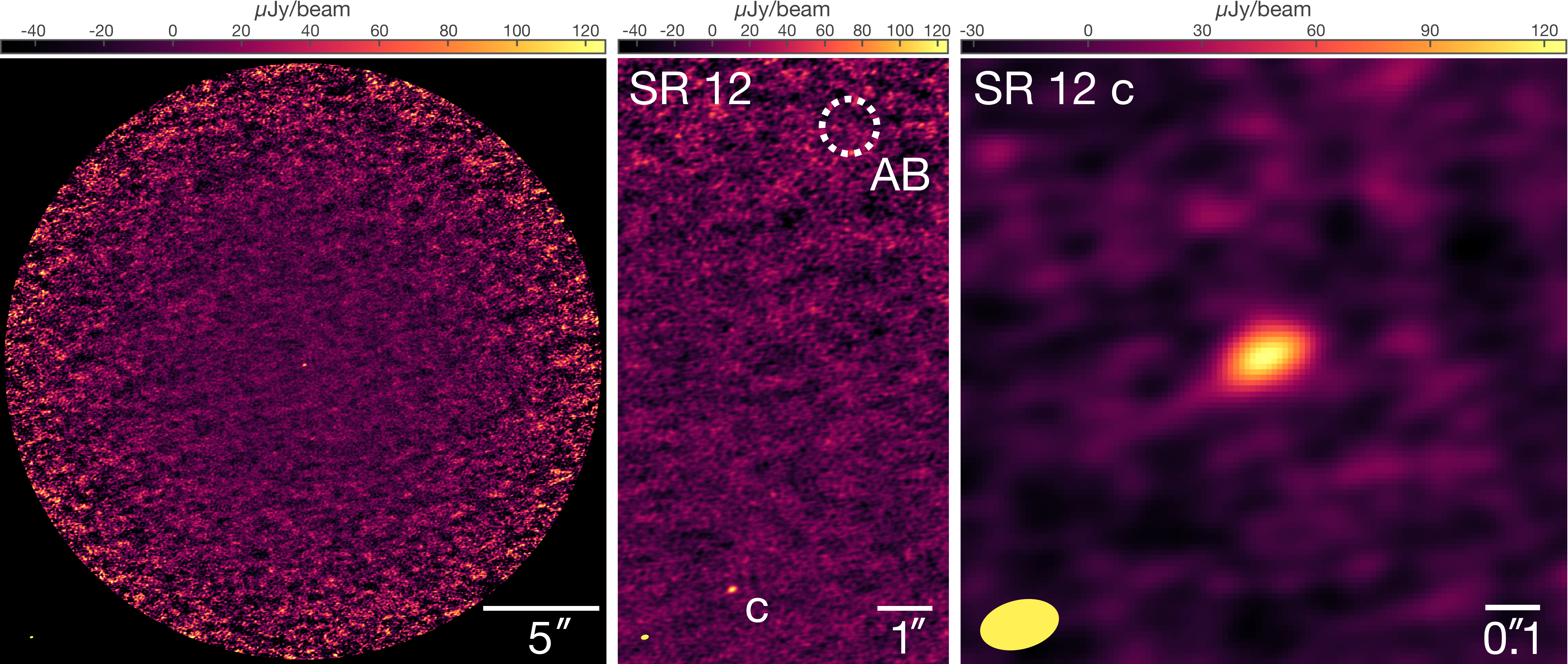}
\caption{ALMA Band 7 discovery of the SR 12 c disk. The $0\farcs147\times0\farcs090$ synthesized beam is plotted at the bottom-left conner of each panel. North is up and east is left. The left panel shows the image where the primary-beam response is $\geqslant$20\% ($12\farcs8$ in radius); the tiny dot at the center is the dust disk around SR 12 c. The middle panel shows the positions of the host binary and the companion. The right panel shows the zoomed-in view of the SR 12 c disk.}
\label{fig:source}
\end{figure*}

Circumplanetary disks (CPDs) around giant exoplanets provide a window into the early solar system when planets and their regular satellites (e.g., Galilean moons) were still growing. The bulk properties and composition of CPDs serve as the initial conditions of planet and satellite formation. Detecting CPDs, however, is challenging even with the unprecedented sensitivity of the Atacama Large Millimeter/submillimeter Array (ALMA). \cite{Andrews21} scrutinized some of the best-resolved protoplanetary disks but did not find any robust candidates. At the moment, the disk around the giant planet PDS 70 c is the only CPD discovered by ALMA \citep{Isella19,B21}. 

Over the past few years, efforts have also been made to search for radio emission associated with the young long-period planetary-mass companions (PMCs) that were discovered in direct-imaging campaigns. PMCs typically have masses around the planet--brown dwarf boundary ($\sim$10--20 $M_{\rm Jup}$) and reside on wide orbits of $\gtrsim$100 au (e.g., 2M1207 b, \citealt{C04}; DH Tau b, \citealt{I05}; 1RXS 1609 b, \citealt{L08}; CT Cha b, \citealt{S08}; FU Tau b, \citealt{L09}; GSC 6214--210 b, \citealt{I11}). While PMC disks may not truly resemble CPDs due to the lack of mass transfer from the circumstellar disks around the host stars (e.g., \citealt{M82,D03}), they can in principle offer a clear view into the satellite-forming environments and mechanisms,  thanks to their ultrawide separations. Moreover, PMC disks appear to be quite common at young ages as \cite{Bowler17} inferred a high disk frequency of $\sim$50\%. Common disk signatures at optical and infrared wavelengths include hydrogen lines, continuum excess, and polarization (e.g., \citealt{Bowler11,Bowler14,Z14,W17b,SM18,S21,vH21,MK21}). When invoking the tidal truncation argument (e.g., \citealt{QT98,AB09}) to estimate PMC disk sizes, which gives an expected disk size of $\sim$$1/3$ Hill radius, this results in a large radius of $\sim$5--60 au for 10--20 $M_{\rm Jup}$ PMCs at 100--1000 au from a 1 $M_\odot$ host star. Nonetheless, up until now, no PMC disks have been successfully imaged with ALMA (e.g., \citealt{Bowler15,M17,R17,W17a,Perez19a,W20}). So far, the most extensive ALMA searches of disks around wide PMCs have been the Band 6 (1.3 mm) survey by \cite{W17a} and the Band 7 (0.88 mm) survey by \cite{W20}, with mean rms noise levels of 40 $\mu$Jy beam$^{-1}$ and 50 $\mu$Jy beam$^{-1}$, respectively. 

These ``snapshot'' observations showed that the dust residing in PMC disks could be compact and optically thick (e.g., \citealt{W17a, Rab19}). Alternatively, another interpretation of the upper limits is that PMC disks may be deficient in dust due to efficient radial drift such that their gas-to-dust ratios are unusually high \citep{Z18}. Deep ALMA imaging is thus essential to detect PMC disks and examine their physical properties.

Here we report the discovery of 0.88 mm continuum emission at the position of SR 12 c. With an rms of 10 $\mu$Jy beam$^{-1}$, $\sim$5 times deeper than the previous 0.88 mm survey in \cite{W20}, this is the most sensitive ALMA imaging of a PMC to date, and the first submillimeter detection of a circumplanetary disk around a widely separated PMC.

\begin{deluxetable*}{@{}ccccc@{}}[t]
\tablewidth{\linewidth}
\tablecaption{ALMA Band 7 Observations of SR 12 \label{tb:obs}}
\tablehead{
\colhead{Start$-$End Time (UTC)} &
\colhead{$N_{\rm ant}$} &
\colhead{Baseline (m)} &
\colhead{$T_{\rm on-source}$ (s)} &
\colhead{Calibrators (Phase, Bandpass, Flux)}  
}
\startdata 
2021-07-08 05:20--06:41	& 43 & 28--3638  & 2601	& J1647$-$2912, J1924$-$2914, J1924$-$2914 \\
2021-07-09 03:07--04:26	& 43 & 28--3396  & 2601	& J1647$-$2912, J1517$-$2422, J1517$-$2422 \\
2021-07-09 04:31--05:52	& 44 & 28--3396  & 2600	& J1647$-$2912, J1924$-$2914, J1924$-$2914 \\
2021-07-10 23:30--23:53	& 46 & 28--3638  & 302	& J1647$-$2912, J1337$-$1257, J1337$-$1257 \\ 
2021-07-11 02:39--03:59	& 45 & 28--3638  & 2601	& J1647$-$2912, J1517$-$2422, J1517$-$2422
\enddata
\end{deluxetable*}

\section{SR 12}
Located in the $\rho$ Ophiuchi star-forming region, SR 12 AB (2MASS J16271951$-$2441403; ROXs 21; YLW 13A) is a weak-line T Tauri binary \citep{Simon87,GR05}, with discrepant spectral-type and mass estimates in the literature. \cite{BA92} suggested that AB is a K4$/$M2.5 binary, while \cite{GR05} fit two blackbody curves of 3428 K and 2500 K to the spectral energy distribution (SED), corresponding to M3 and M8 \citep{PM13}. The separation between AB is 0\farcs21 \citep{K11}, or about 24 au given the Gaia DR2\footnote{SR 12 AB and c have no Gaia EDR3 parallaxes and proper motions \citep{GaiaEDR3}.} distance of $112.5^{+5.8}_{-5.3}$ pc \citep{GaiaDR2,BJ18}. 

The substellar companion SR 12 c (UGCS J162719.66$-$244148.8) at $8\farcs7$ (980 au) from the binary was discovered by \cite{K11}, who also presented 10 yr astrometric measurements demonstrating a common proper motion. It is one of the widest known PMCs and is also the first PMC imaged around a binary star. Given the triangular {\it H}-band spectrum and a low visual extinction of $A_V=$ 1.2--1.7 \citep{K11,Bowler14,SM18}, the age of SR 12 c is generally taken as 2 Myr, which is the median age of the $\rho$ Oph low-extinction stars \citep{W05}. \cite{K11} estimated the mass of SR 12 c to be $0.013\pm0.007~M_\sun$ ($14^{+7}_{-8}~M_{\rm Jup}$) by comparing its age and luminosity to the models of \cite{B97} and \cite{C00}. Using the Gaia DR2 distance, the {\textit J}-band bolometric correction in \cite{Filippazzo15}, and the models in \cite{Baraffe15}, we find a slightly lower bolometric luminosity of log$(L/L_{\odot})=-2.95\pm0.11$ dex and a mass of $11\pm3~M_{\rm Jup}$. The SED fitting in \cite{SM18} yields a spectral type of L$0\pm1$, log {\it g} $=4\pm0.5$, and an effective temperature of $2600\pm100$ K, so the radius of SR 12 c using the Stefan--Boltzmann law is expected to be about 1.6 $R_{\rm Jup}$.

Several lines of evidence indicate that SR 12 c is surrounded by an accretion disk. X-shooter spectroscopy in \cite{SM18,SM19} reveals multiple hydrogen lines, in particular the prominent H$\alpha$ emission, indicating that SR 12 c is accreting at a rate of $10^{-11.08\pm0.40}~M_\sun$ yr$^{-1}$. Thermal dust emission detected with the Spitzer$/$Infrared Array Camera has been reported in \cite{A10}, \cite{G14}, and \cite{MK21}. These disk markers,\footnote{\cite{B08} detected a $\sim$10\% {\it H}-band polarization from a source to the south of SR 12, which they named YLW 13A S. There is no astrometric information for this object, but it is possible that this polarized signal results from the aligned grains in the SR 12 c disk.} together with the ultrawide orbit, make SR 12 c a prime target for ALMA deep imaging.

\section{Observations and Data Reduction}
The ALMA Band 7 observations of SR 12 were carried out on UT 2021 July 8--11, with three 2 GHz windows centered at 333.791, 335.741, and 347.741 GHz, and one 0.938 GHz window centered at 345.787 GHz to image ${}^{12}$CO (3--2) at a resolution of 0.488 MHz. The phase center is the Gaia DR2 position of SR 12 c at $\alpha=16^{\mathrm{h}}27^{\mathrm{m}}19\fs659$ and $\delta=-24\degr41\arcmin49\farcs22$ (epoch J2015.5). The total on-source time is 10,705 s (2.97 hr). Table \ref{tb:obs} lists the start$/$end time, number of antennae, baseline lengths, and calibrators.

Calibrated visibilities were generated using the pipeline 2021.2.0.128 with CASA 6.2.1.7 \citep{McMullin07}. Self-calibration was not performed as the continuum detection (Figure \ref{fig:source}) did not have a sufficiently high signal-to-noise ratio (S$/$N). Diffuse CO emission from the ambient molecular clouds is seen at 0.5--8 \kms~in the LSRK velocity frame, consistent with the channel maps in \cite{deGeus90}, but there is no compact emission directly associated with any of the SR 12 components. The rms in the CO-free channels of 1 \kms~width is $\sim$0.7 mJy beam$^{-1}$. We hence flagged channels that have diffuse CO emission and merged the rest of the window with the three 2 GHz windows. We adopted natural weighting for a better S$/$N and used the CASA task \texttt{tclean} to generate the final continuum map (Figure \ref{fig:source}), which has an rms of 10 $\mu$Jy beam$^{-1}$ around the map center. The size and the position angle of the synthesized beam are $0\farcs147\times0\farcs090$ and 104$\fdg$0.

\begin{figure}[h]
\centering
\figurenum{2}
\includegraphics[width=\columnwidth]{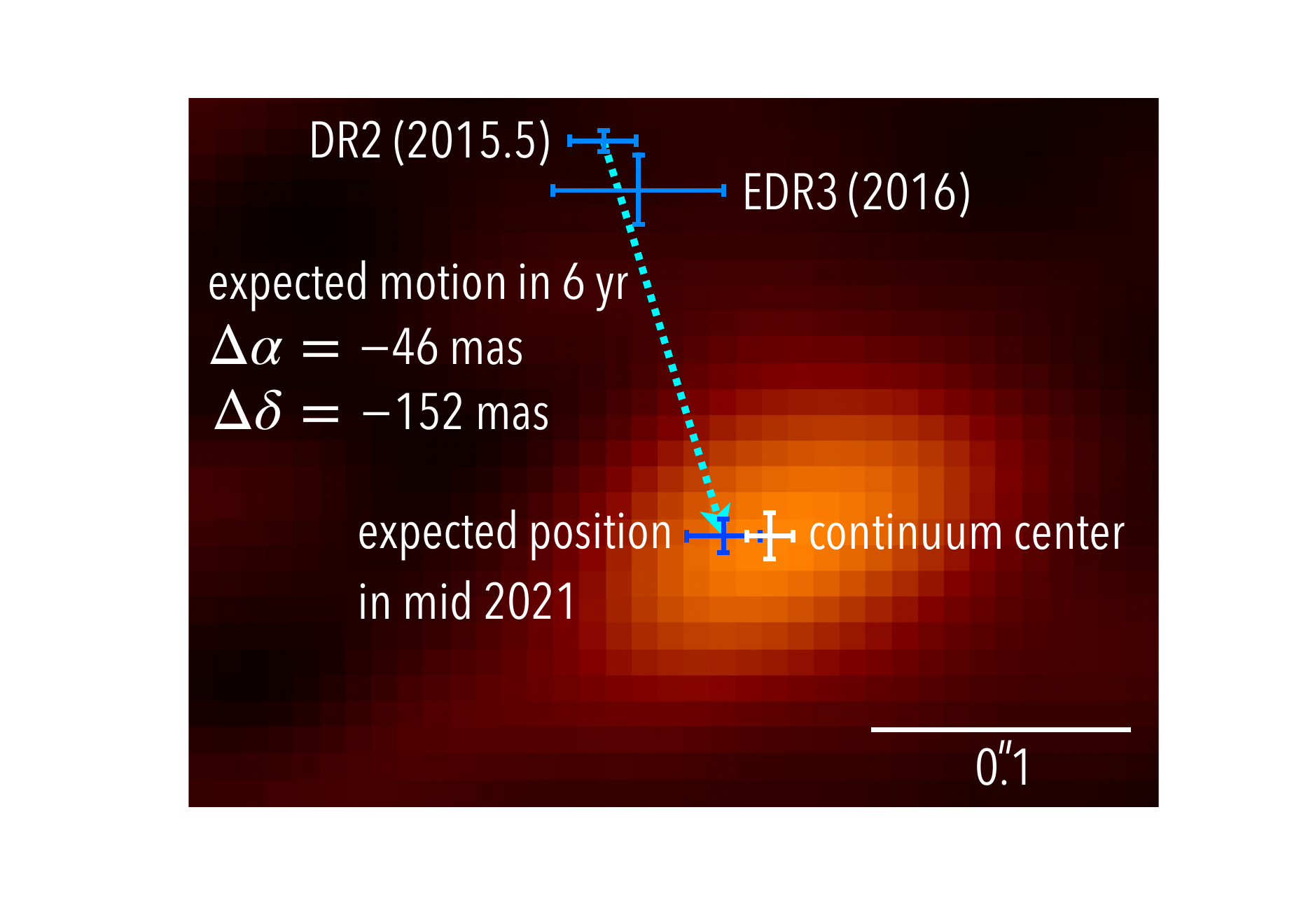}
\caption{Gaia positions of SR 12 c and the ALMA continuum source. The DR2 uncertainties for $\alpha$ and $\delta$ retrieved from the Gaia archive are 13.7 mas and 5.1 mas, respectively. For EDR3, the positional uncertainties are 33.8 mas for $\alpha$ and 14.4 mas for $\delta$. The cyan arrow points to the expected position of c in mid-2021 by adding the 6 yr proper motion of AB to the DR2 position of c. We note that SR 12 c has no DR2$/$EDR3 proper motion available. The J2021.5 positional uncertainties of SR 12 c (15.2 mas for $\alpha$ and 7.1 mas for $\delta$) are calculated by adding the uncertainties of the DR2 position and the 6 yr proper motion in quadrature. The ALMA continuum source has a 10 mas centroiding rms, as shown with the white cross.}
\label{fig:astrometry}
\end{figure}

\section{Results}
\subsection{The SR 12 c Dust Disk}
Figure \ref{fig:source} shows the 0.88 mm continuum map of the SR 12 system. An unresolved object appears close to the phase center, with an ICRS coordinate of $\alpha=16^{\mathrm{h}}27^{\mathrm{m}}19\fs654$ and $\delta=-24\degr41\arcmin49\farcs37$ derived from elliptical Gaussian fitting in the image plane. Visibility fitting with the CASA task \texttt{uvmodelfit} yields the same position. This is slightly different from the positions of SR 12 c in Gaia DR2 ($16^{\mathrm{h}}27^{\mathrm{m}}19\fs659$, $-24\degr41\arcmin49\farcs22$; epoch J2015.5) and EDR3 ($16^{\mathrm{h}}27^{\mathrm{m}}19\fs658$, $-24\degr41\arcmin49\farcs24$; epoch J2016), but as demonstrated in Figure \ref{fig:astrometry}, after accounting for the Gaia DR2 proper motion of the host binary ($\mu_{\alpha^\ast} = -7.7\pm1.1$ mas yr$^{-1}$, $\mu_\delta = -25.3\pm0.8$ mas yr$^{-1}$), the expected position of SR 12 c at the time of our observations matches the position of this 0.88 mm source.

We note that ALMA astrometry should be very accurate. The S$/$N-limited positional accuracy, given by the ALMA Technical Handbook, is beam$_{\rm FWHP}$$/$(S$/$N)$/$0.9, where beam$_{\rm FWHP}$ is the full-width half-power resolution. For our observations, this positional rms is 10 mas. 

It is possible that a background submillimeter galaxy could happen to lie near the position of SR 12 c. We calculate the probability of the point source being a distant galaxy by fitting a power law to the 850$/$870~\micron~differential number counts in \cite{C14}. Integrating from 0.1 mJy to infinity, we find an expectation value of four galaxies brighter than 0.1 mJy within the $\geqslant$20\% primary-beam response (left panel of Figure \ref{fig:source}). The probability of one galaxy being within 0\farcs1 from SR 12 c is 0.02\%. Thus, this 0.88 mm source is most likely a dust disk around SR 12 c.

\subsection{Dust Disk Radius}
\label{sect:dustradius}
As the dust disk appears unresolved,\footnote{The disk is not resolved with Briggs weighting (robust $=$ 0.5) in \texttt{tclean} either, where the beam size is $0\farcs108\times0\farcs066$.} its radius is likely to be smaller than 5 au. We assume a point source model in the CASA task {\tt uvmodelfit} and obtain a flux density of $127\pm14~\mu$Jy, where the uncertainty includes a 10\% flux calibration error. This is consistent with the peak flux density of 126 $\mu$Jy directly measured from the image, indicating that the disk is indeed not resolved. It is about 1.5 times brighter than the CPD around PDS 70 c (86 $\mu$Jy; \citealt{B21}) given their very similar distances (112.5 pc to SR 12 and 112.4 pc to PDS 70).

Because optically thick emission scales with surface area, we can apply this assumption to estimate the disk radius needed to account for the observed emission. Such an assumption may be expected for PMC disks as the simulations in \cite{Rab19} show that PMC disks are optically thick at millimeter wavelengths. In brief, we substitute a radial profile of the disk temperature $T(r)$ into the Planck function $B_\nu$ and integrate $B_\nu(T(r))$ from the disk inner radius $r_{\rm in}$ to the outer radius $r_{\rm out}$ to calculate the flux density $F_\nu$, i.e., $F_\nu = \frac{2\pi \langle\cos i \rangle}{D^2}\int^{r_{\rm out}}_{r_{\rm in}} B_\nu(T(r))r\,dr$, where $D$ is the distance to the source and $i$ is the disk inclination whose mean value is $\langle \cos i \rangle=2/\pi$. As $F_\nu$ is mostly sensitive to $r_{\rm out}$, we can find a disk extent compatible with the observed flux density. This outer radius, however, may be a lower limit because if dust emission is optically thin and scales with the dust mass, then an extended disk can have a similar brightness as long as there is enough dust.

We follow \cite{Isella14,Isella19} to calculate the temperature profile as $T(r)^4 = T_{\rm \ast,\,irr}^4 + T_{\rm PMC,\,irr}^4 + T^4_{\rm acc}$, where $T_{\rm \ast,\,irr}$ and $T_{\rm PMC,\,irr}$ accounts for the irradiation from the star and the PMC, respectively, and $T_{\rm acc}$ comes from viscous heating due to accretion. The functional form of each term can be found in the original articles. At 980 au, we can safely ignore the irradiation from the host binary. After substituting the bolometric luminosity of SR 12 c into $T_{\rm PMC,\,irr}$ and the mass, radius, and accretion rate into $T_{\rm acc}$, we find a disk outer radius of 0.34 au ($\sim$720 $R_{\rm Jup}$), much smaller than the minimum radius that can be resolved by the beam ($\sim$5 au). We note that in this analysis, the disk inner radius is taken to be the PMC radius because the disk temperature there is still below the dust sublimation temperature of $\sim$1500 K.\footnote{The sublimation temperature depends on a variety of factors such as grain composition and gas density (e.g., \citealt{Kobayashi11}). However, given the small area of the high-temperature disk regions, our results remain valid even if we exclude inner regions that would be above 1000 K.} Although the inferred radius depends on the assumed temperature profile, our ALMA observations suggest that the dust disk is probably very small.

\subsection{Dust Mass}
\label{sect:dustmass}
Applying the optically thin assumption to what is actually optically thick emission may significantly underestimate the true dust mass (e.g., \citealt{BE19}). But to compare SR 12 c with PDS 70 c, we first find the mean disk temperature and then adopt the opacity assumptions in \cite{B21} to calculate the dust mass. Integrating $T(r) = (T_{\rm PMC,\,irr}^4 + T^4_{\rm acc})^{0.25}$ from 1.6 $R_{\rm Jup}$ to 5 au (the upper limit of the SR 12 c disk radius), we find a mean disk temperature of $T_{\rm mean}=24$ K. This result is not very sensitive to the accretion rate because PMC irradiation dominates the disk temperature profile. Assuming the disk only contains 1 mm grains with an opacity of $\kappa_{\rm 1\,mm}=3.63$ cm$^2$ g$^{-1}$ \citep{Birnstiel18}, we substitute $T_{\rm mean}$ and $\kappa_{\rm 1\,mm}$ into $M_{\rm dust} = \frac{F_\nu D^2}{\kappa B_\nu(T_{\rm mean})}$ \citep{H83} and find a dust mass of $0.012~M_{\oplus}$ (0.95 $M_{\rm Moon}$), two times more massive than the PDS 70 c disk, 0.006 $M_{\oplus}$ (0.47 $M_{\rm Moon}$).\footnote{We find a mean disk temperature of 29 K for the PDS 70 c disk when integrating $T(r)$ to 1 au, the disk outer radius constrained by \cite{B21}. This is slightly higher than the temperature (26 K) used by \cite{B21} to calculate the dust mass. Therefore, our mass estimates are slightly lower than theirs (0.007 $M_{\oplus}$ for 1 mm grains and 0.031 $M_{\oplus}$ for 1 \micron~grains).}

\cite{Z18} found that CPDs can quickly lose their millimeter-sized dust in just hundreds of years due to very efficient radial drift. To be detectable at (sub)millimeter wavelengths, CPDs need to be constantly replenished, have a high surface density and$/$or substructures to slow down the radial drift, or contain enough micron-sized grains that can emit substantial millimeter emission. SR 12 c is not seen in $^{12}$CO (3--2) and unlikely to be replenished from the surroundings, so its disk may contain mostly micron-sized grains or it has substructures possibly carved by satellites. If the millimeter-sized dust grains are already depleted such that only 1 \micron~grains remain, we find a dust mass of $0.054~M_{\oplus}$ (4.4 $M_{\rm Moon}$) assuming an opacity of 0.79 cm$^2$ g$^{-1}$ \citep{Birnstiel18}, again two times higher than the PDS 70 c CPD, 0.027 $M_{\oplus}$ (2.2 $M_{\rm Moon}$).

Another planetary-mass object whose disk has been detected by ALMA is OTS 44 (2MASS J11100934$-$7632178), a free-floating $\sim$12 $M_{\rm Jup}$ object accreting at $7.6\times10^{-12}~M_\sun$ yr$^{-1}$ \citep{J13}. Its disk has a 1.3 mm flux density of 101 $\mu$Jy \citep{B17}, translating to 238 $\mu$Jy at 0.88 mm assuming a spectral index of 2.2 \citep{Ribas17}. Adopting a Gaia DR2 distance of 192 pc to the Chamaeleon I star-forming region \citep{Dzib18} and assuming the same $T_{\rm mean}$ and opacity as the SR 12 c disk,\footnote{There is no tighter constraint on the OTS 44 disk size to calculate the mean disk temperature (the beam size in \cite{B17} is $1\farcs6 \times 1\farcs6$).} we find a dust mass of 0.064 $M_\oplus$ (5.2 $M_{\rm Moon}$) with $\kappa_{\rm 1\, mm}$ and 0.295 $M_\oplus$ (24 $M_{\rm Moon}$) with $\kappa_{\rm 1\, \mu m}$.

\begin{figure}[h]
\centering
\figurenum{3}
\includegraphics[width=0.992\columnwidth]{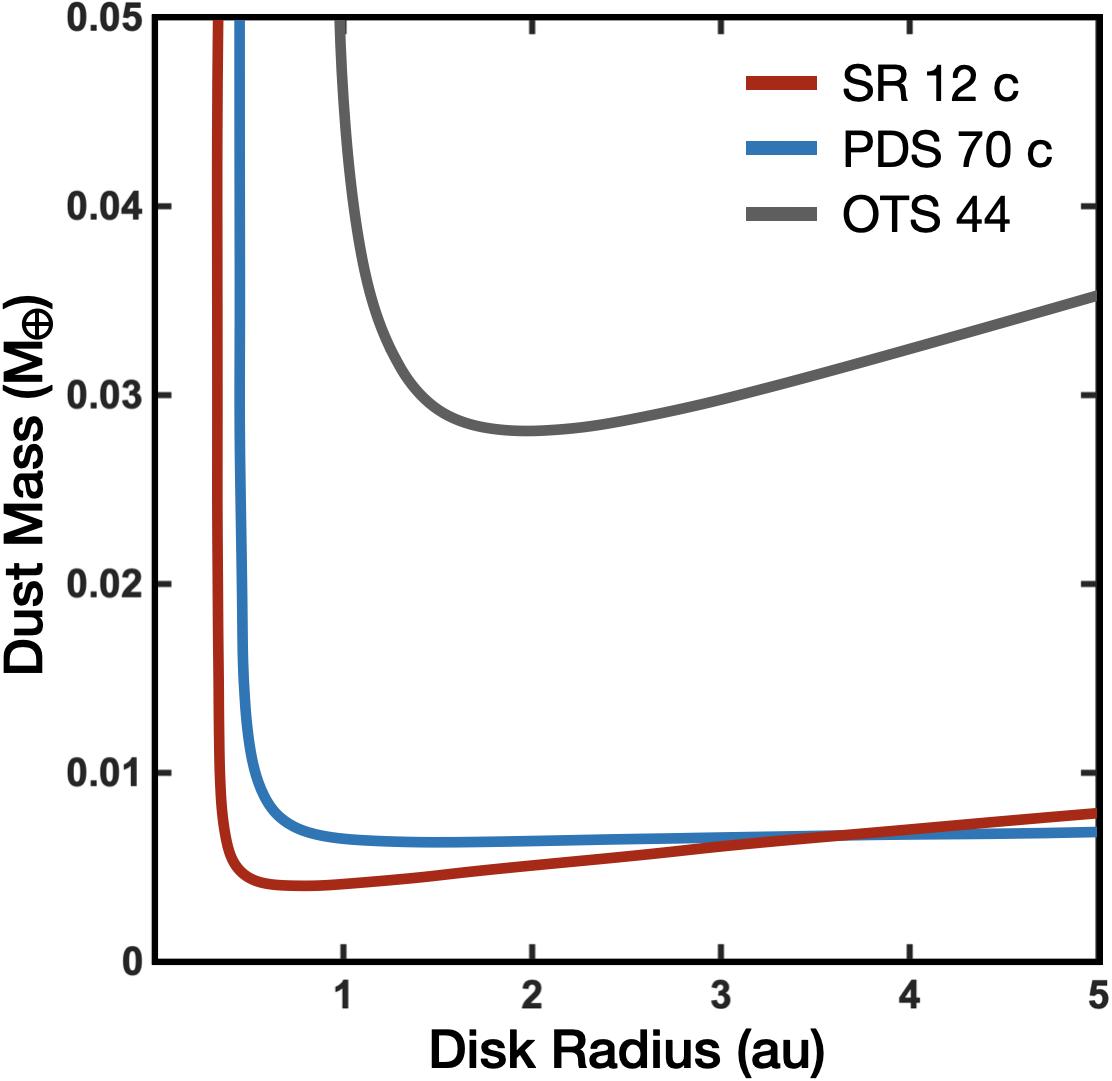}
\caption{Dust masses vs. disk outer radii for a surface density profile $\propto$ $1/r$ and $\kappa_{\rm 1\,mm}=3.63$ cm$^2$ g$^{-1}$. In the optically thin regime, $M_{\rm SR\,12\,c}$ and $M_{\rm PDS\,70\,c}$ are $\sim$0.007 $M_\oplus$, and $M_{\rm OTS\,44}$ is $\sim$0.03 $M_\oplus$. The curves for $\kappa_{\rm 1\,\mu m}=0.79$ cm$^2$ g$^{-1}$ (not shown) are four to five times larger in values: $M_{\rm SR\,12\,c}$ and $M_{\rm PDS\,70\,c}$ are $\sim$0.03 $M_\oplus$, and $M_{\rm OTS\,44}$ is $\sim$0.15 $M_\oplus$.}
\label{fig:profile}
\end{figure}

The dust-mass estimates above are likely upper limits because the mean temperature will increase when the disk size decreases. A more comprehensive approach to infer the dust mass and trace the transition of the optical depth is to fit a surface density model to match the observed flux density, as demonstrated by \cite{Isella14,Isella19}. We follow their prescription by assuming a surface density $\Sigma(r)\propto 1/r$ to calculate the expected continuum flux density with  Equation (4) in \cite{Isella14}. The dust mass is derived from $\int^{r_{\rm out}}_{r_{\rm in}}\Sigma(r)2\pi r dr$, where the disk inner radius $r_{\rm in}$ equals the source radius because, as has been found previously, the disk temperature does not exceed 1500 K in the innermost region. Figure \ref{fig:profile} shows the dust mass as a function of the disk outer radius for an opacity of $\kappa_{\rm 1\,mm}=3.63$ cm$^2$ g$^{-1}$. When the disks are larger than $\sim$1 au, their dust masses are roughly independent of the outer radii, indicating an optically thin dust emission. The SR 12 c disk and the PDS 70 c CPD may have comparable dust masses, but $\sim$5 times less than the OTS 44 disk---a substantial spread in circumplanetary dust content at young ages. On the other hand, when the disks are smaller than $\sim$1 au, they quickly become optically thick such that the curves turn vertical.

While at face value, these disks seem to be low in dust content, they may still have the potential to form massive satellites like the Galilean moons (total mass of 0.066 $M_\oplus$) during the final stage of gas accretion if they are indeed compact, dense, and thus have dust masses much higher than the optically thin values. Compared with OTS 44 and SR 12 c, the continuous replenishment of solids from the circumstellar disk may further facilitate massive satellite formation around PDS 70 c (e.g., \citealt{CW02}).

\subsection{Gas-to-dust Ratio and Small Gas Disk}
The accretion rate of SR 12 c measured by \cite{SM19}, $10^{-11.08\pm0.40}~M_\sun$ yr$^{-1}$ ($\sim$$9\times10^{-9}~M_{\rm Jup}$ yr$^{-1}$),\footnote{The actual accretion rate may be higher based on the new accretion shock models in \cite{Aoyama21}.} implies a substantial amount of gas within the disk: on the order of 0.03 $M_{\rm Jup}$, assuming that accretion at this rate can last for another 3 Myr before the disk may dissipate (e.g., \citealt{H01}). With the dust masses derived in Figure \ref{fig:profile}, 0.007 $M_\oplus$ for 1 mm dust and 0.03 $M_\oplus$ for 1~\micron~dust, and the 0.40 dex uncertainty in the accretion rate, we find gas-to-dust ratios in the ranges of 470--3000 and 110--700, respectively. The derived ratios appear to be higher than the interstellar medium value of 100 and maybe much higher as disks around lower-mass objects may survive longer (e.g., \citealt{LM12}). However, as the disk probably contains more dust than computed for the optically thin assumption and the accretion rate may be highly variable, it remains unclear if the actual ratio is indeed inconsistent with the canonical value.
 
As ${}^{12}$CO emission is usually optically thick throughout the disk, the line flux strongly depends on the disk area but weakly correlates with the disk mass. For instance, \cite{Miotello21} show that for a compact disk around a very-low-mass star, ${}^{12}$CO (3--2) only increases by a factor of $\sim$1.5--2 when the disk mass increases by a factor of 1000. A similar trend is present for the PMC disk models in \cite{Rab19}. Assuming the ${}^{12}$CO (3--2) emission from SR 12 c is not absorbed by foreground clouds, our nondetection favors a small gas disk. In the simulations of \cite{Rab19}, a gas disk with a radius of $\sim$10 au can have a peak line strength of $\sim$4--5 mJy beam$^{-1}$ per 1 \kms~channel when scaled to the distance to SR 12 c (see their Figure 5), which would be detectable at $\gtrsim$5$\sigma$ by our observations (the beam size in their simulations, $0\farcs1\times0\farcs1$, is comparable to our observations). Therefore, the gas disk of SR 12 c unlikely extends to 1$/$3 Hill radius ($\sim$50 au) but instead may be smaller than 10 au. 

Compact gas disks may in fact be fairly common from the stellar- to planetary-mass regimes. In dense regions such as the Orion Nebula Cluster, most disks are likely truncated by photoevaporation or dynamical encounters with other stars (e.g., \citealt{E08,BE20}), while in low-density star-forming regions such as Lupus, other internal mechanisms may counteract the viscous spreading to create a high fraction of compact disks \citep{Miotello21}. Although the majority of PMCs are located in less dense regions (e.g., Chamaeleon, Taurus, Lupus, Upper Scorpius, etc.), it remains to be investigated whether external or internal processes could influence the extent of PMC disks.

The lack of CO emission may also result from CO being frozen out onto grains. However, compact disks are warmer, so CO should more easily sublimate off the grains. Indeed, the dust temperature calculation in Section \ref{sect:dustradius} shows that most parts of the disk are warmer than 20 K, the temperature at which most CO desorbs into the gas phase \citep{Bergin95}. Therefore, CO depletion may not play an important role in the SR 12 c disk.

\subsection{Circumbinary Material around SR 12 AB} 
Dust continuum from the host binary is not detected down to an rms of 19 $\mu$Jy beam$^{-1}$, which is consistent with results from \cite{MK21} that the binary has no Spitzer infrared excess. Circumbinary material may have mostly dispersed due to the orbital motion between the stars, and we find a 3$\sigma$ dust-mass limit of $<$0.007 $M_{\oplus}$ ($<$0.56 $M_{\rm Moon}$) assuming identical $\kappa_{\rm 1\,mm}$ and $T_{\rm mean}$ $=$ 20 K.

\subsection{Scaling Relation}
Our discovery indicates that more PMC disks may be accessible in comparatively deep ALMA observations. Indeed, a $\sim$0.1 mJy disk could resemble a 2$\sigma$--3$\sigma$ noise blob in the previous 0.88 mm survey \citep{W20}, so it may be fruitful to revisit accreting PMCs with sensitivities comparable to our observations. As disk properties may reflect the satellite-forming processes and timescales, it is also important to investigate if the disks around planetary-mass objects, either bound or isolated, follow the scaling relationships of circumstellar disks. Compared with Figure 4 in \cite{W20}, both the PDS 70 c and SR 12 c disks are in line with the relationship between the disk flux density and the host mass in the stellar regime. More disk detections, or tighter constraints, are needed to examine if young PMC disks are systematically fainter than what would be expected from the stellar relationship, as suggested in \cite{W20}.

\section{Summary}
Our new ALMA 0.88 mm imaging made the first submillimeter detection of an accretion disk around a wide-orbit PMC. The SR 12 c dust disk has a flux density of 127 $\mu$Jy, appears unresolved by the $\sim$$0\farcs1$ beam, and may have a small radius of $\sim$0.3 au if the continuum is optically thick. The gas disk may have a high gas-to-dust ratio of $\gg$100 and be very compact as well. A comparison of the optically thin dust masses between PDS 70 c, SR 12 c, and OTS 44 implies a wide range of dust content among young planetary-mass objects.

\begin{acknowledgements}
We thank the reviewer for constructive comments. Y.-L.W. is grateful for the support from the Ministry of Science and Technology and the Ministry of Education. J.A.E. acknowledges support from NSF awards 1745406 and 1811290. This paper makes use of the following ALMA data: ADS$/$JAO.ALMA $\#$2019.1.01212.S. ALMA is a partnership of ESO (representing its member states), NSF (USA) and NINS (Japan), together with NRC (Canada), MOST and ASIAA (Taiwan), and KASI (Republic of Korea), in cooperation with the Republic of Chile. The Joint ALMA Observatory is operated by ESO, AUI$/$NRAO and NAOJ. This research has made use of MATLAB version R2020b.
\end{acknowledgements}

\facilities{ALMA.}
\software{CASA \citep{McMullin07}.}


\begin{thebibliography}{}
\bibitem[Alves de Oliveira et al.(2010)]{A10} Alves de Oliveira, C., Moraux, E., Bouvier, J., et al.\ 2010, \aap, 515, A75. doi:10.1051/0004-6361/200913900
\bibitem[Andrews et al.(2021)]{Andrews21} Andrews, S.~M., Elder, W., Zhang, S., et al.\ 2021, \apj, 916, 51. doi:10.3847/1538-4357/ac00b9
\bibitem[Ayliffe \& Bate(2009)]{AB09} Ayliffe, B.~A. \& Bate, M.~R.\ 2009, \mnras, 397, 657. doi:10.1111/j.1365-2966.2009.15002.x
\bibitem[Aoyama et al.(2021)]{Aoyama21} Aoyama, Y., Marleau, G.-D., Ikoma, M., et al.\ 2021, \apjl, 917, L30. doi:10.3847/2041-8213/ac19bd
\bibitem[Bailer-Jones et al.(2018)]{BJ18} Bailer-Jones, C.~A.~L., Rybizki, J., Fouesneau, M., et al.\ 2018, \aj, 156, 58. doi:10.3847/1538-3881/aacb21
\bibitem[Ballering \& Eisner(2019)]{BE19} Ballering, N.~P. \& Eisner, J.~A.\ 2019, \aj, 157, 144. doi:10.3847/1538-3881/ab0a56
\bibitem[Baraffe et al.(2015)]{Baraffe15} Baraffe, I., Homeier, D., Allard, F., et al.\ 2015, \aap, 577, A42. doi:10.1051/0004-6361/201425481
\bibitem[Bayo et al.(2017)]{B17} Bayo, A., Joergens, V., Liu, Y., et al.\ 2017, \apjl, 841, L11. doi:10.3847/2041-8213/aa7046
\bibitem[Beckford et al.(2008)]{B08} Beckford, A.~F., Lucas, P.~W., Chrysostomou, A.~C., et al.\ 2008, \mnras, 384, 907. doi:10.1111/j.1365-2966.2007.12715.x
\bibitem[Benisty et al.(2021)]{B21} Benisty, M., Bae, J., Facchini, S., et al.\ 2021, \apjl, 916, L2. doi:10.3847/2041-8213/ac0f83
\bibitem[Bergin et al.(1995)]{Bergin95} Bergin, E.~A., Langer, W.~D., \& Goldsmith, P.~F.\ 1995, \apj, 441, 222. doi:10.1086/175351
\bibitem[Birnstiel et al.(2018)]{Birnstiel18} Birnstiel, T., Dullemond, C.~P., Zhu, Z., et al.\ 2018, \apjl, 869, L45. doi:10.3847/2041-8213/aaf743
\bibitem[Bouvier \& Appenzeller(1992)]{BA92} Bouvier, J. \& Appenzeller, I.\ 1992, \aaps, 92, 481
\bibitem[Bowler et al.(2015)]{Bowler15} Bowler, B.~P., Andrews, S.~M., Kraus, A.~L., et al.\ 2015, \apjl, 805, L17. doi:10.1088/2041-8205/805/2/L17
\bibitem[Bowler et al.(2017)]{Bowler17} Bowler, B.~P., Kraus, A.~L., Bryan, M.~L., et al.\ 2017, \aj, 154, 165. doi:10.3847/1538-3881/aa88bd
\bibitem[Bowler et al.(2011)]{Bowler11} Bowler, B.~P., Liu, M.~C., Kraus, A.~L., et al.\ 2011, \apj, 743, 148. doi:10.1088/0004-637X/743/2/148
\bibitem[Bowler et al.(2014)]{Bowler14} Bowler, B.~P., Liu, M.~C., Kraus, A.~L., et al.\ 2014, \apj, 784, 65. doi:10.1088/0004-637X/784/1/65
\bibitem[Boyden \& Eisner(2020)]{BE20} Boyden, R.~D. \& Eisner, J.~A.\ 2020, \apj, 894, 74. doi:10.3847/1538-4357/ab86b7
\bibitem[Burrows et al.(1997)]{B97} Burrows, A., Marley, M., Hubbard, W.~B., et al.\ 1997, \apj, 491, 856. doi:10.1086/305002
\bibitem[Canup \& Ward(2002)]{CW02} Canup, R.~M. \& Ward, W.~R.\ 2002, \aj, 124, 3404. doi:10.1086/344684
\bibitem[Casey et al.(2014)]{C14} Casey, C.~M., Narayanan, D., \& Cooray, A.\ 2014, \physrep, 541, 45. doi:10.1016/j.physrep.2014.02.009
\bibitem[Chabrier et al.(2000)]{C00} Chabrier, G., Baraffe, I., Allard, F., et al.\ 2000, \apj, 542, 464. doi:10.1086/309513
\bibitem[Chauvin et al.(2004)]{C04} Chauvin, G., Lagrange, A.-M., Dumas, C., et al.\ 2004, \aap, 425, L29. doi:10.1051/0004-6361:200400056
\bibitem[D'Angelo et al.(2003)]{D03} D'Angelo, G., Kley, W., \& Henning, T.\ 2003, \apj, 586, 540. doi:10.1086/367555
\bibitem[de Geus et al.(1990)]{deGeus90} de Geus, E.~J., Bronfman, L., \& Thaddeus, P.\ 1990, \aap, 231, 137
\bibitem[Dzib et al.(2018)]{Dzib18} Dzib, S.~A., Loinard, L., Ortiz-Le{\'o}n, G.~N., et al.\ 2018, \apj, 867, 151. doi:10.3847/1538-4357/aae687
\bibitem[Eisner et al.(2008)]{E08} Eisner, J.~A., Plambeck, R.~L., Carpenter, J.~M., et al.\ 2008, \apj, 683, 304. doi:10.1086/588524
\bibitem[Filippazzo et al.(2015)]{Filippazzo15} Filippazzo, J.~C., Rice, E.~L., Faherty, J., et al.\ 2015, \apj, 810, 158. doi:10.1088/0004-637X/810/2/158
\bibitem[Gaia Collaboration et al.(2018)]{GaiaDR2} Gaia Collaboration, Brown, A.~G.~A., Vallenari, A., et al.\ 2018, \aap, 616, A1. doi:10.1051/0004-6361/201833051
\bibitem[Gaia Collaboration et al.(2021)]{GaiaEDR3} Gaia Collaboration, Brown, A.~G.~A., Vallenari, A., et al.\ 2021, \aap, 649, A1. doi:10.1051/0004-6361/202039657
\bibitem[Gras-Vel{\'a}zquez \& Ray(2005)]{GR05} Gras-Vel{\'a}zquez, {\`A}. \& Ray, T.~P.\ 2005, \aap, 443, 541. doi:10.1051/0004-6361:20042397
\bibitem[G{\"u}nther et al.(2014)]{G14} G{\"u}nther, H.~M., Cody, A.~M., Covey, K.~R., et al.\ 2014, \aj, 148, 122. doi:10.1088/0004-6256/148/6/122
\bibitem[Haisch et al.(2001)]{H01} Haisch, K.~E., Lada, E.~A., \& Lada, C.~J.\ 2001, \apjl, 553, L153. doi:10.1086/320685
\bibitem[Hildebrand(1983)]{H83} Hildebrand, R.~H.\ 1983, \qjras, 24, 267
\bibitem[Ireland et al.(2011)]{I11} Ireland, M.~J., Kraus, A., Martinache, F., et al.\ 2011, \apj, 726, 113. doi:10.1088/0004-637X/726/2/113
\bibitem[Isella et al.(2019)]{Isella19} Isella, A., Benisty, M., Teague, R., et al.\ 2019, \apjl, 879, L25. doi:10.3847/2041-8213/ab2a12
\bibitem[Isella et al.(2014)]{Isella14} Isella, A., Chandler, C.~J., Carpenter, J.~M., et al.\ 2014, \apj, 788, 129. doi:10.1088/0004-637X/788/2/129
\bibitem[Itoh et al.(2005)]{I05} Itoh, Y., Hayashi, M., Tamura, M., et al.\ 2005, \apj, 620, 984. doi:10.1086/427086
\bibitem[Joergens et al.(2013)]{J13} Joergens, V., Bonnefoy, M., Liu, Y., et al.\ 2013, \aap, 558, L7. doi:10.1051/0004-6361/201322432
\bibitem[Kobayashi et al.(2011)]{Kobayashi11} Kobayashi, H., Kimura, H., Watanabe, S.-. i ., et al.\ 2011, EP\&S, 63, 1067. doi:10.5047/eps.2011.03.012
\bibitem[Kuzuhara et al.(2011)]{K11} Kuzuhara, M., Tamura, M., Ishii, M., et al.\ 2011, \aj, 141, 119. doi:10.1088/0004-6256/141/4/119
\bibitem[Lafreni{\`e}re et al.(2008)]{L08} Lafreni{\`e}re, D., Jayawardhana, R., \& van Kerkwijk, M.~H.\ 2008, \apjl, 689, L153. doi:10.1086/595870
\bibitem[Luhman \& Mamajek(2012)]{LM12} Luhman, K.~L. \& Mamajek, E.~E.\ 2012, \apj, 758, 31. doi:10.1088/0004-637X/758/1/31
\bibitem[Luhman et al.(2009)]{L09} Luhman, K.~L., Mamajek, E.~E., Allen, P.~R., et al.\ 2009, \apj, 691, 1265. doi:10.1088/0004-637X/691/2/1265
\bibitem[MacGregor et al.(2017)]{M17} MacGregor, M.~A., Wilner, D.~J., Czekala, I., et al.\ 2017, \apj, 835, 17. doi:10.3847/1538-4357/835/1/17
\bibitem[Martinez \& Kraus(2022)]{MK21} Martinez, R.~A. \& Kraus, A.~L.\ 2022, \aj, 163, 36. doi:10.3847/1538-3881/ac3745
\bibitem[McMullin et al.(2007)]{McMullin07} McMullin, J.~P., Waters, B., Schiebel, D., et al.\ 2007, adass XVI, 376, 127
\bibitem[Miki(1982)]{M82} Miki, S.\ 1982, Progress of Theoretical Physics, 67, 1053. doi:10.1143/PTP.67.1053
\bibitem[Miotello et al.(2021)]{Miotello21} Miotello, A., Rosotti, G., Ansdell, M., et al.\ 2021, \aap, 651, A48. doi:10.1051/0004-6361/202140550
\bibitem[Pecaut \& Mamajek(2013)]{PM13} Pecaut, M.~J. \& Mamajek, E.~E.\ 2013, \apjs, 208, 9. doi:10.1088/0067-0049/208/1/9
\bibitem[P{\'e}rez et al.(2019)]{Perez19a} P{\'e}rez, S., Marino, S., Casassus, S., et al.\ 2019, \mnras, 488, 1005. doi:10.1093/mnras/stz1775
\bibitem[Quillen \& Trilling(1998)]{QT98} Quillen, A.~C. \& Trilling, D.~E.\ 1998, \apj, 508, 707. doi:10.1086/306421
\bibitem[Rab et al.(2019)]{Rab19} Rab, C., Kamp, I., Ginski, C., et al.\ 2019, \aap, 624, A16. doi:10.1051/0004-6361/201834899
\bibitem[Ribas et al.(2017)]{Ribas17} Ribas, {\'A}., Espaillat, C.~C., Mac{\'\i}as, E., et al.\ 2017, \apj, 849, 63. doi:10.3847/1538-4357/aa8e99
\bibitem[Ricci et al.(2017)]{R17} Ricci, L., Cazzoletti, P., Czekala, I., et al.\ 2017, \aj, 154, 24. doi:10.3847/1538-3881/aa78a0
\bibitem[Santamar{\'\i}a-Miranda et al.(2018)]{SM18} Santamar{\'\i}a-Miranda, A., C{\'a}ceres, C., Schreiber, M.~R., et al.\ 2018, \mnras, 475, 2994. doi:10.1093/mnras/stx3325
\bibitem[Santamar{\'\i}a-Miranda et al.(2019)]{SM19} Santamar{\'\i}a-Miranda, A., C{\'a}ceres, C., Schreiber, M.~R., et al.\ 2019, \mnras, 488, 5852. doi:10.1093/mnras/stz2173
\bibitem[Schmidt et al.(2008)]{S08} Schmidt, T.~O.~B., Neuh{\"a}user, R., Seifahrt, A., et al.\ 2008, \aap, 491, 311. doi:10.1051/0004-6361:20078840
\bibitem[Simon et al.(1987)]{Simon87} Simon, M., Howell, R.~R., Longmore, A.~J., et al.\ 1987, \apj, 320, 344. doi:10.1086/165548
\bibitem[Stolker et al.(2021)]{S21} Stolker, T., Haffert, S.~Y., Kesseli, A.~Y., et al.\ 2021, \aj, 162, 286. doi:10.3847/1538-3881/ac2c7f
\bibitem[van Holstein et al.(2021)]{vH21} van Holstein, R.~G., Stolker, T., Jensen-Clem, R., et al.\ 2021, \aap, 647, A21. doi:10.1051/0004-6361/202039290
\bibitem[Wilking et al.(2005)]{W05} Wilking, B.~A., Meyer, M.~R., Robinson, J.~G., et al.\ 2005, \aj, 130, 1733. doi:10.1086/432758
\bibitem[Wu et al.(2020)]{W20} Wu, Y.-L., Bowler, B.~P., Sheehan, P.~D., et al.\ 2020, \aj, 159, 229. doi:10.3847/1538-3881/ab818c
\bibitem[Wu et al.(2017a)]{W17a} Wu, Y.-L., Close, L.~M., Eisner, J.~A., et al.\ 2017a, \aj, 154, 234. doi:10.3847/1538-3881/aa93db
\bibitem[Wu et al.(2017b)]{W17b} Wu, Y.-L., Sheehan, P.~D., Males, J.~R., et al.\ 2017b, \apj, 836, 223. doi:10.3847/1538-4357/aa5b96
\bibitem[Zhou et al.(2014)]{Z14} Zhou, Y., Herczeg, G.~J., Kraus, A.~L., et al.\ 2014, \apjl, 783, L17. doi:10.1088/2041-8205/783/1/L17
\bibitem[Zhu et al.(2018)]{Z18} Zhu, Z., Andrews, S.~M., \& Isella, A.\ 2018, \mnras, 479, 1850. doi:10.1093/mnras/sty1503
\end{thebibliography}
\end{document}